\documentclass{jfm}
\usepackage{graphicx}
\usepackage{epstopdf, epsfig}
\usepackage{color}
\definecolor{AliceBlue}{rgb}{0.94,0.97,1.00}
\definecolor{AntiqueWhite1}{rgb}{1.00,0.94,0.86}
\definecolor{AntiqueWhite2}{rgb}{0.93,0.87,0.80}
\definecolor{AntiqueWhite3}{rgb}{0.80,0.75,0.69}
\definecolor{AntiqueWhite4}{rgb}{0.55,0.51,0.47}
\definecolor{AntiqueWhite}{rgb}{0.98,0.92,0.84}
\definecolor{BlanchedAlmond}{rgb}{1.00,0.92,0.80}
\definecolor{BlueViolet}{rgb}{0.54,0.17,0.89}
\definecolor{CadetBlue1}{rgb}{0.60,0.96,1.00}
\definecolor{CadetBlue2}{rgb}{0.56,0.90,0.93}
\definecolor{CadetBlue3}{rgb}{0.48,0.77,0.80}
\definecolor{CadetBlue4}{rgb}{0.33,0.53,0.55}
\definecolor{CadetBlue}{rgb}{0.37,0.62,0.63}
\definecolor{CornflowerBlue}{rgb}{0.39,0.58,0.93}
\definecolor{DarkBlue}{rgb}{0.00,0.00,0.55}
\definecolor{DarkCyan}{rgb}{0.00,0.55,0.55}
\definecolor{DarkGoldenrod1}{rgb}{1.00,0.73,0.06}
\definecolor{DarkGoldenrod2}{rgb}{0.93,0.68,0.05}
\definecolor{DarkGoldenrod3}{rgb}{0.80,0.58,0.05}
\definecolor{DarkGoldenrod4}{rgb}{0.55,0.40,0.03}
\definecolor{DarkGoldenrod}{rgb}{0.72,0.53,0.04}
\definecolor{DarkGray}{rgb}{0.66,0.66,0.66}
\definecolor{DarkGreen}{rgb}{0.00,0.39,0.00}
\definecolor{DarkGrey}{rgb}{0.66,0.66,0.66}
\definecolor{DarkKhaki}{rgb}{0.74,0.72,0.42}
\definecolor{DarkMagenta}{rgb}{0.55,0.00,0.55}
\definecolor{DarkOliveGreen1}{rgb}{0.79,1.00,0.44}
\definecolor{DarkOliveGreen2}{rgb}{0.74,0.93,0.41}
\definecolor{DarkOliveGreen3}{rgb}{0.64,0.80,0.35}
\definecolor{DarkOliveGreen4}{rgb}{0.43,0.55,0.24}
\definecolor{DarkOliveGreen}{rgb}{0.33,0.42,0.18}
\definecolor{DarkOrange1}{rgb}{1.00,0.50,0.00}
\definecolor{DarkOrange2}{rgb}{0.93,0.46,0.00}
\definecolor{DarkOrange3}{rgb}{0.80,0.40,0.00}
\definecolor{DarkOrange4}{rgb}{0.55,0.27,0.00}
\definecolor{DarkOrange}{rgb}{1.00,0.55,0.00}
\definecolor{DarkOrchid1}{rgb}{0.75,0.24,1.00}
\definecolor{DarkOrchid2}{rgb}{0.70,0.23,0.93}
\definecolor{DarkOrchid3}{rgb}{0.60,0.20,0.80}
\definecolor{DarkOrchid4}{rgb}{0.41,0.13,0.55}
\definecolor{DarkOrchid}{rgb}{0.60,0.20,0.80}
\definecolor{DarkRed}{rgb}{0.55,0.00,0.00}
\definecolor{DarkSalmon}{rgb}{0.91,0.59,0.48}
\definecolor{DarkSeaGreen1}{rgb}{0.76,1.00,0.76}
\definecolor{DarkSeaGreen2}{rgb}{0.71,0.93,0.71}
\definecolor{DarkSeaGreen3}{rgb}{0.61,0.80,0.61}
\definecolor{DarkSeaGreen4}{rgb}{0.41,0.55,0.41}
\definecolor{DarkSeaGreen}{rgb}{0.56,0.74,0.56}
\definecolor{DarkSlateBlue}{rgb}{0.28,0.24,0.55}
\definecolor{DarkSlateGray1}{rgb}{0.59,1.00,1.00}
\definecolor{DarkSlateGray2}{rgb}{0.55,0.93,0.93}
\definecolor{DarkSlateGray3}{rgb}{0.47,0.80,0.80}
\definecolor{DarkSlateGray4}{rgb}{0.32,0.55,0.55}
\definecolor{DarkSlateGray}{rgb}{0.18,0.31,0.31}
\definecolor{DarkSlateGrey}{rgb}{0.18,0.31,0.31}
\definecolor{DarkTurquoise}{rgb}{0.00,0.81,0.82}
\definecolor{DarkViolet}{rgb}{0.58,0.00,0.83}
\definecolor{DeepPink1}{rgb}{1.00,0.08,0.58}
\definecolor{DeepPink2}{rgb}{0.93,0.07,0.54}
\definecolor{DeepPink3}{rgb}{0.80,0.06,0.46}
\definecolor{DeepPink4}{rgb}{0.55,0.04,0.31}
\definecolor{DeepPink}{rgb}{1.00,0.08,0.58}
\definecolor{DeepSkyBlue1}{rgb}{0.00,0.75,1.00}
\definecolor{DeepSkyBlue2}{rgb}{0.00,0.70,0.93}
\definecolor{DeepSkyBlue3}{rgb}{0.00,0.60,0.80}
\definecolor{DeepSkyBlue4}{rgb}{0.00,0.41,0.55}
\definecolor{DeepSkyBlue}{rgb}{0.00,0.75,1.00}
\definecolor{DimGray}{rgb}{0.41,0.41,0.41}
\definecolor{DimGrey}{rgb}{0.41,0.41,0.41}
\definecolor{DodgerBlue1}{rgb}{0.12,0.56,1.00}
\definecolor{DodgerBlue2}{rgb}{0.11,0.53,0.93}
\definecolor{DodgerBlue3}{rgb}{0.09,0.45,0.80}
\definecolor{DodgerBlue4}{rgb}{0.06,0.31,0.55}
\definecolor{DodgerBlue}{rgb}{0.12,0.56,1.00}
\definecolor{FloralWhite}{rgb}{1.00,0.98,0.94}
\definecolor{ForestGreen}{rgb}{0.13,0.55,0.13}
\definecolor{GhostWhite}{rgb}{0.97,0.97,1.00}
\definecolor{GreenYellow}{rgb}{0.68,1.00,0.18}
\definecolor{HotPink1}{rgb}{1.00,0.43,0.71}
\definecolor{HotPink2}{rgb}{0.93,0.42,0.65}
\definecolor{HotPink3}{rgb}{0.80,0.38,0.56}
\definecolor{HotPink4}{rgb}{0.55,0.23,0.38}
\definecolor{HotPink}{rgb}{1.00,0.41,0.71}
\definecolor{IndianRed1}{rgb}{1.00,0.42,0.42}
\definecolor{IndianRed2}{rgb}{0.93,0.39,0.39}
\definecolor{IndianRed3}{rgb}{0.80,0.33,0.33}
\definecolor{IndianRed4}{rgb}{0.55,0.23,0.23}
\definecolor{IndianRed}{rgb}{0.80,0.36,0.36}
\definecolor{LavenderBlush1}{rgb}{1.00,0.94,0.96}
\definecolor{LavenderBlush2}{rgb}{0.93,0.88,0.90}
\definecolor{LavenderBlush3}{rgb}{0.80,0.76,0.77}
\definecolor{LavenderBlush4}{rgb}{0.55,0.51,0.53}
\definecolor{LavenderBlush}{rgb}{1.00,0.94,0.96}
\definecolor{LawnGreen}{rgb}{0.49,0.99,0.00}
\definecolor{LemonChiffon1}{rgb}{1.00,0.98,0.80}
\definecolor{LemonChiffon2}{rgb}{0.93,0.91,0.75}
\definecolor{LemonChiffon3}{rgb}{0.80,0.79,0.65}
\definecolor{LemonChiffon4}{rgb}{0.55,0.54,0.44}
\definecolor{LemonChiffon}{rgb}{1.00,0.98,0.80}
\definecolor{LightBlue1}{rgb}{0.75,0.94,1.00}
\definecolor{LightBlue2}{rgb}{0.70,0.87,0.93}
\definecolor{LightBlue3}{rgb}{0.60,0.75,0.80}
\definecolor{LightBlue4}{rgb}{0.41,0.51,0.55}
\definecolor{LightBlue}{rgb}{0.68,0.85,0.90}
\definecolor{LightCoral}{rgb}{0.94,0.50,0.50}
\definecolor{LightCyan1}{rgb}{0.88,1.00,1.00}
\definecolor{LightCyan2}{rgb}{0.82,0.93,0.93}
\definecolor{LightCyan3}{rgb}{0.71,0.80,0.80}
\definecolor{LightCyan4}{rgb}{0.48,0.55,0.55}
\definecolor{LightCyan}{rgb}{0.88,1.00,1.00}
\definecolor{LightGoldenrod1}{rgb}{1.00,0.93,0.55}
\definecolor{LightGoldenrod2}{rgb}{0.93,0.86,0.51}
\definecolor{LightGoldenrod3}{rgb}{0.80,0.75,0.44}
\definecolor{LightGoldenrod4}{rgb}{0.55,0.51,0.30}
\definecolor{LightGoldenrodYellow}{rgb}{0.98,0.98,0.82}
\definecolor{LightGoldenrod}{rgb}{0.93,0.87,0.51}
\definecolor{LightGray}{rgb}{0.83,0.83,0.83}
\definecolor{LightGreen}{rgb}{0.56,0.93,0.56}
\definecolor{LightGrey}{rgb}{0.83,0.83,0.83}
\definecolor{LightPink1}{rgb}{1.00,0.68,0.73}
\definecolor{LightPink2}{rgb}{0.93,0.64,0.68}
\definecolor{LightPink3}{rgb}{0.80,0.55,0.58}
\definecolor{LightPink4}{rgb}{0.55,0.37,0.40}
\definecolor{LightPink}{rgb}{1.00,0.71,0.76}
\definecolor{LightSalmon1}{rgb}{1.00,0.63,0.48}
\definecolor{LightSalmon2}{rgb}{0.93,0.58,0.45}
\definecolor{LightSalmon3}{rgb}{0.80,0.51,0.38}
\definecolor{LightSalmon4}{rgb}{0.55,0.34,0.26}
\definecolor{LightSalmon}{rgb}{1.00,0.63,0.48}
\definecolor{LightSeaGreen}{rgb}{0.13,0.70,0.67}
\definecolor{LightSkyBlue1}{rgb}{0.69,0.89,1.00}
\definecolor{LightSkyBlue2}{rgb}{0.64,0.83,0.93}
\definecolor{LightSkyBlue3}{rgb}{0.55,0.71,0.80}
\definecolor{LightSkyBlue4}{rgb}{0.38,0.48,0.55}
\definecolor{LightSkyBlue}{rgb}{0.53,0.81,0.98}
\definecolor{LightSlateBlue}{rgb}{0.52,0.44,1.00}
\definecolor{LightSlateGray}{rgb}{0.47,0.53,0.60}
\definecolor{LightSlateGrey}{rgb}{0.47,0.53,0.60}
\definecolor{LightSteelBlue1}{rgb}{0.79,0.88,1.00}
\definecolor{LightSteelBlue2}{rgb}{0.74,0.82,0.93}
\definecolor{LightSteelBlue3}{rgb}{0.64,0.71,0.80}
\definecolor{LightSteelBlue4}{rgb}{0.43,0.48,0.55}
\definecolor{LightSteelBlue}{rgb}{0.69,0.77,0.87}
\definecolor{LightYellow1}{rgb}{1.00,1.00,0.88}
\definecolor{LightYellow2}{rgb}{0.93,0.93,0.82}
\definecolor{LightYellow3}{rgb}{0.80,0.80,0.71}
\definecolor{LightYellow4}{rgb}{0.55,0.55,0.48}
\definecolor{LightYellow}{rgb}{1.00,1.00,0.88}
\definecolor{LimeGreen}{rgb}{0.20,0.80,0.20}
\definecolor{MediumAquamarine}{rgb}{0.40,0.80,0.67}
\definecolor{MediumBlue}{rgb}{0.00,0.00,0.80}
\definecolor{MediumOrchid1}{rgb}{0.88,0.40,1.00}
\definecolor{MediumOrchid2}{rgb}{0.82,0.37,0.93}
\definecolor{MediumOrchid3}{rgb}{0.71,0.32,0.80}
\definecolor{MediumOrchid4}{rgb}{0.48,0.22,0.55}
\definecolor{MediumOrchid}{rgb}{0.73,0.33,0.83}
\definecolor{MediumPurple1}{rgb}{0.67,0.51,1.00}
\definecolor{MediumPurple2}{rgb}{0.62,0.47,0.93}
\definecolor{MediumPurple3}{rgb}{0.54,0.41,0.80}
\definecolor{MediumPurple4}{rgb}{0.36,0.28,0.55}
\definecolor{MediumPurple}{rgb}{0.58,0.44,0.86}
\definecolor{MediumSeaGreen}{rgb}{0.24,0.70,0.44}
\definecolor{MediumSlateBlue}{rgb}{0.48,0.41,0.93}
\definecolor{MediumSpringGreen}{rgb}{0.00,0.98,0.60}
\definecolor{MediumTurquoise}{rgb}{0.28,0.82,0.80}
\definecolor{MediumVioletRed}{rgb}{0.78,0.08,0.52}
\definecolor{MidnightBlue}{rgb}{0.10,0.10,0.44}
\definecolor{MintCream}{rgb}{0.96,1.00,0.98}
\definecolor{MistyRose1}{rgb}{1.00,0.89,0.88}
\definecolor{MistyRose2}{rgb}{0.93,0.84,0.82}
\definecolor{MistyRose3}{rgb}{0.80,0.72,0.71}
\definecolor{MistyRose4}{rgb}{0.55,0.49,0.48}
\definecolor{MistyRose}{rgb}{1.00,0.89,0.88}
\definecolor{NavajoWhite1}{rgb}{1.00,0.87,0.68}
\definecolor{NavajoWhite2}{rgb}{0.93,0.81,0.63}
\definecolor{NavajoWhite3}{rgb}{0.80,0.70,0.55}
\definecolor{NavajoWhite4}{rgb}{0.55,0.47,0.37}
\definecolor{NavajoWhite}{rgb}{1.00,0.87,0.68}
\definecolor{NavyBlue}{rgb}{0.00,0.00,0.50}
\definecolor{OldLace}{rgb}{0.99,0.96,0.90}
\definecolor{OliveDrab1}{rgb}{0.75,1.00,0.24}
\definecolor{OliveDrab2}{rgb}{0.70,0.93,0.23}
\definecolor{OliveDrab3}{rgb}{0.60,0.80,0.20}
\definecolor{OliveDrab4}{rgb}{0.41,0.55,0.13}
\definecolor{OliveDrab}{rgb}{0.42,0.56,0.14}
\definecolor{OrangeRed1}{rgb}{1.00,0.27,0.00}
\definecolor{OrangeRed2}{rgb}{0.93,0.25,0.00}
\definecolor{OrangeRed3}{rgb}{0.80,0.22,0.00}
\definecolor{OrangeRed4}{rgb}{0.55,0.15,0.00}
\definecolor{OrangeRed}{rgb}{1.00,0.27,0.00}
\definecolor{PaleGoldenrod}{rgb}{0.93,0.91,0.67}
\definecolor{PaleGreen1}{rgb}{0.60,1.00,0.60}
\definecolor{PaleGreen2}{rgb}{0.56,0.93,0.56}
\definecolor{PaleGreen3}{rgb}{0.49,0.80,0.49}
\definecolor{PaleGreen4}{rgb}{0.33,0.55,0.33}
\definecolor{PaleGreen}{rgb}{0.60,0.98,0.60}
\definecolor{PaleTurquoise1}{rgb}{0.73,1.00,1.00}
\definecolor{PaleTurquoise2}{rgb}{0.68,0.93,0.93}
\definecolor{PaleTurquoise3}{rgb}{0.59,0.80,0.80}
\definecolor{PaleTurquoise4}{rgb}{0.40,0.55,0.55}
\definecolor{PaleTurquoise}{rgb}{0.69,0.93,0.93}
\definecolor{PaleVioletRed1}{rgb}{1.00,0.51,0.67}
\definecolor{PaleVioletRed2}{rgb}{0.93,0.47,0.62}
\definecolor{PaleVioletRed3}{rgb}{0.80,0.41,0.54}
\definecolor{PaleVioletRed4}{rgb}{0.55,0.28,0.36}
\definecolor{PaleVioletRed}{rgb}{0.86,0.44,0.58}
\definecolor{PapayaWhip}{rgb}{1.00,0.94,0.84}
\definecolor{PeachPuff1}{rgb}{1.00,0.85,0.73}
\definecolor{PeachPuff2}{rgb}{0.93,0.80,0.68}
\definecolor{PeachPuff3}{rgb}{0.80,0.69,0.58}
\definecolor{PeachPuff4}{rgb}{0.55,0.47,0.40}
\definecolor{PeachPuff}{rgb}{1.00,0.85,0.73}
\definecolor{PowderBlue}{rgb}{0.69,0.88,0.90}
\definecolor{RosyBrown1}{rgb}{1.00,0.76,0.76}
\definecolor{RosyBrown2}{rgb}{0.93,0.71,0.71}
\definecolor{RosyBrown3}{rgb}{0.80,0.61,0.61}
\definecolor{RosyBrown4}{rgb}{0.55,0.41,0.41}
\definecolor{RosyBrown}{rgb}{0.74,0.56,0.56}
\definecolor{RoyalBlue1}{rgb}{0.28,0.46,1.00}
\definecolor{RoyalBlue2}{rgb}{0.26,0.43,0.93}
\definecolor{RoyalBlue3}{rgb}{0.23,0.37,0.80}
\definecolor{RoyalBlue4}{rgb}{0.15,0.25,0.55}
\definecolor{RoyalBlue}{rgb}{0.25,0.41,0.88}
\definecolor{SaddleBrown}{rgb}{0.55,0.27,0.07}
\definecolor{SandyBrown}{rgb}{0.96,0.64,0.38}
\definecolor{SeaGreen1}{rgb}{0.33,1.00,0.62}
\definecolor{SeaGreen2}{rgb}{0.31,0.93,0.58}
\definecolor{SeaGreen3}{rgb}{0.26,0.80,0.50}
\definecolor{SeaGreen4}{rgb}{0.18,0.55,0.34}
\definecolor{SeaGreen}{rgb}{0.18,0.55,0.34}
\definecolor{SkyBlue1}{rgb}{0.53,0.81,1.00}
\definecolor{SkyBlue2}{rgb}{0.49,0.75,0.93}
\definecolor{SkyBlue3}{rgb}{0.42,0.65,0.80}
\definecolor{SkyBlue4}{rgb}{0.29,0.44,0.55}
\definecolor{SkyBlue}{rgb}{0.53,0.81,0.92}
\definecolor{SlateBlue1}{rgb}{0.51,0.44,1.00}
\definecolor{SlateBlue2}{rgb}{0.48,0.40,0.93}
\definecolor{SlateBlue3}{rgb}{0.41,0.35,0.80}
\definecolor{SlateBlue4}{rgb}{0.28,0.24,0.55}
\definecolor{SlateBlue}{rgb}{0.42,0.35,0.80}
\definecolor{SlateGray1}{rgb}{0.78,0.89,1.00}
\definecolor{SlateGray2}{rgb}{0.73,0.83,0.93}
\definecolor{SlateGray3}{rgb}{0.62,0.71,0.80}
\definecolor{SlateGray4}{rgb}{0.42,0.48,0.55}
\definecolor{SlateGray}{rgb}{0.44,0.50,0.56}
\definecolor{SlateGrey}{rgb}{0.44,0.50,0.56}
\definecolor{SpringGreen1}{rgb}{0.00,1.00,0.50}
\definecolor{SpringGreen2}{rgb}{0.00,0.93,0.46}
\definecolor{SpringGreen3}{rgb}{0.00,0.80,0.40}
\definecolor{SpringGreen4}{rgb}{0.00,0.55,0.27}
\definecolor{SpringGreen}{rgb}{0.00,1.00,0.50}
\definecolor{SteelBlue1}{rgb}{0.39,0.72,1.00}
\definecolor{SteelBlue2}{rgb}{0.36,0.67,0.93}
\definecolor{SteelBlue3}{rgb}{0.31,0.58,0.80}
\definecolor{SteelBlue4}{rgb}{0.21,0.39,0.55}
\definecolor{SteelBlue}{rgb}{0.27,0.51,0.71}
\definecolor{VioletRed1}{rgb}{1.00,0.24,0.59}
\definecolor{VioletRed2}{rgb}{0.93,0.23,0.55}
\definecolor{VioletRed3}{rgb}{0.80,0.20,0.47}
\definecolor{VioletRed4}{rgb}{0.55,0.13,0.32}
\definecolor{VioletRed}{rgb}{0.82,0.13,0.56}
\definecolor{WhiteSmoke}{rgb}{0.96,0.96,0.96}
\definecolor{YellowGreen}{rgb}{0.60,0.80,0.20}
\definecolor{aliceblue}{rgb}{0.94,0.97,1.00}
\definecolor{antiquewhite}{rgb}{0.98,0.92,0.84}
\definecolor{aquamarine1}{rgb}{0.50,1.00,0.83}
\definecolor{aquamarine2}{rgb}{0.46,0.93,0.78}
\definecolor{aquamarine3}{rgb}{0.40,0.80,0.67}
\definecolor{aquamarine4}{rgb}{0.27,0.55,0.45}
\definecolor{aquamarine}{rgb}{0.50,1.00,0.83}
\definecolor{azure1}{rgb}{0.94,1.00,1.00}
\definecolor{azure2}{rgb}{0.88,0.93,0.93}
\definecolor{azure3}{rgb}{0.76,0.80,0.80}
\definecolor{azure4}{rgb}{0.51,0.55,0.55}
\definecolor{azure}{rgb}{0.94,1.00,1.00}
\definecolor{beige}{rgb}{0.96,0.96,0.86}
\definecolor{bisque1}{rgb}{1.00,0.89,0.77}
\definecolor{bisque2}{rgb}{0.93,0.84,0.72}
\definecolor{bisque3}{rgb}{0.80,0.72,0.62}
\definecolor{bisque4}{rgb}{0.55,0.49,0.42}
\definecolor{bisque}{rgb}{1.00,0.89,0.77}
\definecolor{black}{rgb}{0.00,0.00,0.00}
\definecolor{blanchedalmond}{rgb}{1.00,0.92,0.80}
\definecolor{blue1}{rgb}{0.00,0.00,1.00}
\definecolor{blue2}{rgb}{0.00,0.00,0.93}
\definecolor{blue3}{rgb}{0.00,0.00,0.80}
\definecolor{blue4}{rgb}{0.00,0.00,0.55}
\definecolor{blueviolet}{rgb}{0.54,0.17,0.89}
\definecolor{blue}{rgb}{0.00,0.00,1.00}
\definecolor{brown1}{rgb}{1.00,0.25,0.25}
\definecolor{brown2}{rgb}{0.93,0.23,0.23}
\definecolor{brown3}{rgb}{0.80,0.20,0.20}
\definecolor{brown4}{rgb}{0.55,0.14,0.14}
\definecolor{brown}{rgb}{0.65,0.16,0.16}
\definecolor{burlywood1}{rgb}{1.00,0.83,0.61}
\definecolor{burlywood2}{rgb}{0.93,0.77,0.57}
\definecolor{burlywood3}{rgb}{0.80,0.67,0.49}
\definecolor{burlywood4}{rgb}{0.55,0.45,0.33}
\definecolor{burlywood}{rgb}{0.87,0.72,0.53}
\definecolor{cadetblue}{rgb}{0.37,0.62,0.63}
\definecolor{chartreuse1}{rgb}{0.50,1.00,0.00}
\definecolor{chartreuse2}{rgb}{0.46,0.93,0.00}
\definecolor{chartreuse3}{rgb}{0.40,0.80,0.00}
\definecolor{chartreuse4}{rgb}{0.27,0.55,0.00}
\definecolor{chartreuse}{rgb}{0.50,1.00,0.00}
\definecolor{chocolate1}{rgb}{1.00,0.50,0.14}
\definecolor{chocolate2}{rgb}{0.93,0.46,0.13}
\definecolor{chocolate3}{rgb}{0.80,0.40,0.11}
\definecolor{chocolate4}{rgb}{0.55,0.27,0.07}
\definecolor{chocolate}{rgb}{0.82,0.41,0.12}
\definecolor{coral1}{rgb}{1.00,0.45,0.34}
\definecolor{coral2}{rgb}{0.93,0.42,0.31}
\definecolor{coral3}{rgb}{0.80,0.36,0.27}
\definecolor{coral4}{rgb}{0.55,0.24,0.18}
\definecolor{coral}{rgb}{1.00,0.50,0.31}
\definecolor{cornflowerblue}{rgb}{0.39,0.58,0.93}
\definecolor{cornsilk1}{rgb}{1.00,0.97,0.86}
\definecolor{cornsilk2}{rgb}{0.93,0.91,0.80}
\definecolor{cornsilk3}{rgb}{0.80,0.78,0.69}
\definecolor{cornsilk4}{rgb}{0.55,0.53,0.47}
\definecolor{cornsilk}{rgb}{1.00,0.97,0.86}
\definecolor{cyan1}{rgb}{0.00,1.00,1.00}
\definecolor{cyan2}{rgb}{0.00,0.93,0.93}
\definecolor{cyan3}{rgb}{0.00,0.80,0.80}
\definecolor{cyan4}{rgb}{0.00,0.55,0.55}
\definecolor{cyan}{rgb}{0.00,1.00,1.00}
\definecolor{darkblue}{rgb}{0.00,0.00,0.55}
\definecolor{darkcyan}{rgb}{0.00,0.55,0.55}
\definecolor{darkgoldenrod}{rgb}{0.72,0.53,0.04}
\definecolor{darkgray}{rgb}{0.66,0.66,0.66}
\definecolor{darkgreen}{rgb}{0.00,0.39,0.00}
\definecolor{darkgrey}{rgb}{0.66,0.66,0.66}
\definecolor{darkkhaki}{rgb}{0.74,0.72,0.42}
\definecolor{darkmagenta}{rgb}{0.55,0.00,0.55}
\definecolor{darkolive}{rgb}{0.33,0.42,0.18}
\definecolor{darkorange}{rgb}{1.00,0.55,0.00}
\definecolor{darkorchid}{rgb}{0.60,0.20,0.80}
\definecolor{darkred}{rgb}{0.55,0.00,0.00}
\definecolor{darksalmon}{rgb}{0.91,0.59,0.48}
\definecolor{darksea}{rgb}{0.56,0.74,0.56}
\definecolor{darkslate}{rgb}{0.18,0.31,0.31}
\definecolor{darkslate}{rgb}{0.18,0.31,0.31}
\definecolor{darkslate}{rgb}{0.28,0.24,0.55}
\definecolor{darkturquoise}{rgb}{0.00,0.81,0.82}
\definecolor{darkviolet}{rgb}{0.58,0.00,0.83}
\definecolor{deeppink}{rgb}{1.00,0.08,0.58}
\definecolor{deepsky}{rgb}{0.00,0.75,1.00}
\definecolor{dimgray}{rgb}{0.41,0.41,0.41}
\definecolor{dimgrey}{rgb}{0.41,0.41,0.41}
\definecolor{dodgerblue}{rgb}{0.12,0.56,1.00}
\definecolor{firebrick1}{rgb}{1.00,0.19,0.19}
\definecolor{firebrick2}{rgb}{0.93,0.17,0.17}
\definecolor{firebrick3}{rgb}{0.80,0.15,0.15}
\definecolor{firebrick4}{rgb}{0.55,0.10,0.10}
\definecolor{firebrick}{rgb}{0.70,0.13,0.13}
\definecolor{floralwhite}{rgb}{1.00,0.98,0.94}
\definecolor{forestgreen}{rgb}{0.13,0.55,0.13}
\definecolor{gainsboro}{rgb}{0.86,0.86,0.86}
\definecolor{ghostwhite}{rgb}{0.97,0.97,1.00}
\definecolor{gold1}{rgb}{1.00,0.84,0.00}
\definecolor{gold2}{rgb}{0.93,0.79,0.00}
\definecolor{gold3}{rgb}{0.80,0.68,0.00}
\definecolor{gold4}{rgb}{0.55,0.46,0.00}
\definecolor{goldenrod1}{rgb}{1.00,0.76,0.15}
\definecolor{goldenrod2}{rgb}{0.93,0.71,0.13}
\definecolor{goldenrod3}{rgb}{0.80,0.61,0.11}
\definecolor{goldenrod4}{rgb}{0.55,0.41,0.08}
\definecolor{goldenrod}{rgb}{0.85,0.65,0.13}
\definecolor{gold}{rgb}{1.00,0.84,0.00}
\definecolor{gray0}{rgb}{0.00,0.00,0.00}
\definecolor{gray100}{rgb}{1.00,1.00,1.00}
\definecolor{gray10}{rgb}{0.10,0.10,0.10}
\definecolor{gray11}{rgb}{0.11,0.11,0.11}
\definecolor{gray12}{rgb}{0.12,0.12,0.12}
\definecolor{gray13}{rgb}{0.13,0.13,0.13}
\definecolor{gray14}{rgb}{0.14,0.14,0.14}
\definecolor{gray15}{rgb}{0.15,0.15,0.15}
\definecolor{gray16}{rgb}{0.16,0.16,0.16}
\definecolor{gray17}{rgb}{0.17,0.17,0.17}
\definecolor{gray18}{rgb}{0.18,0.18,0.18}
\definecolor{gray19}{rgb}{0.19,0.19,0.19}
\definecolor{gray1}{rgb}{0.01,0.01,0.01}
\definecolor{gray20}{rgb}{0.20,0.20,0.20}
\definecolor{gray21}{rgb}{0.21,0.21,0.21}
\definecolor{gray22}{rgb}{0.22,0.22,0.22}
\definecolor{gray23}{rgb}{0.23,0.23,0.23}
\definecolor{gray24}{rgb}{0.24,0.24,0.24}
\definecolor{gray25}{rgb}{0.25,0.25,0.25}
\definecolor{gray26}{rgb}{0.26,0.26,0.26}
\definecolor{gray27}{rgb}{0.27,0.27,0.27}
\definecolor{gray28}{rgb}{0.28,0.28,0.28}
\definecolor{gray29}{rgb}{0.29,0.29,0.29}
\definecolor{gray2}{rgb}{0.02,0.02,0.02}
\definecolor{gray30}{rgb}{0.30,0.30,0.30}
\definecolor{gray31}{rgb}{0.31,0.31,0.31}
\definecolor{gray32}{rgb}{0.32,0.32,0.32}
\definecolor{gray33}{rgb}{0.33,0.33,0.33}
\definecolor{gray34}{rgb}{0.34,0.34,0.34}
\definecolor{gray35}{rgb}{0.35,0.35,0.35}
\definecolor{gray36}{rgb}{0.36,0.36,0.36}
\definecolor{gray37}{rgb}{0.37,0.37,0.37}
\definecolor{gray38}{rgb}{0.38,0.38,0.38}
\definecolor{gray39}{rgb}{0.39,0.39,0.39}
\definecolor{gray3}{rgb}{0.03,0.03,0.03}
\definecolor{gray40}{rgb}{0.40,0.40,0.40}
\definecolor{gray41}{rgb}{0.41,0.41,0.41}
\definecolor{gray42}{rgb}{0.42,0.42,0.42}
\definecolor{gray43}{rgb}{0.43,0.43,0.43}
\definecolor{gray44}{rgb}{0.44,0.44,0.44}
\definecolor{gray45}{rgb}{0.45,0.45,0.45}
\definecolor{gray46}{rgb}{0.46,0.46,0.46}
\definecolor{gray47}{rgb}{0.47,0.47,0.47}
\definecolor{gray48}{rgb}{0.48,0.48,0.48}
\definecolor{gray49}{rgb}{0.49,0.49,0.49}
\definecolor{gray4}{rgb}{0.04,0.04,0.04}
\definecolor{gray50}{rgb}{0.50,0.50,0.50}
\definecolor{gray51}{rgb}{0.51,0.51,0.51}
\definecolor{gray52}{rgb}{0.52,0.52,0.52}
\definecolor{gray53}{rgb}{0.53,0.53,0.53}
\definecolor{gray54}{rgb}{0.54,0.54,0.54}
\definecolor{gray55}{rgb}{0.55,0.55,0.55}
\definecolor{gray56}{rgb}{0.56,0.56,0.56}
\definecolor{gray57}{rgb}{0.57,0.57,0.57}
\definecolor{gray58}{rgb}{0.58,0.58,0.58}
\definecolor{gray59}{rgb}{0.59,0.59,0.59}
\definecolor{gray5}{rgb}{0.05,0.05,0.05}
\definecolor{gray60}{rgb}{0.60,0.60,0.60}
\definecolor{gray61}{rgb}{0.61,0.61,0.61}
\definecolor{gray62}{rgb}{0.62,0.62,0.62}
\definecolor{gray63}{rgb}{0.63,0.63,0.63}
\definecolor{gray64}{rgb}{0.64,0.64,0.64}
\definecolor{gray65}{rgb}{0.65,0.65,0.65}
\definecolor{gray66}{rgb}{0.66,0.66,0.66}
\definecolor{gray67}{rgb}{0.67,0.67,0.67}
\definecolor{gray68}{rgb}{0.68,0.68,0.68}
\definecolor{gray69}{rgb}{0.69,0.69,0.69}
\definecolor{gray6}{rgb}{0.06,0.06,0.06}
\definecolor{gray70}{rgb}{0.70,0.70,0.70}
\definecolor{gray71}{rgb}{0.71,0.71,0.71}
\definecolor{gray72}{rgb}{0.72,0.72,0.72}
\definecolor{gray73}{rgb}{0.73,0.73,0.73}
\definecolor{gray74}{rgb}{0.74,0.74,0.74}
\definecolor{gray75}{rgb}{0.75,0.75,0.75}
\definecolor{gray76}{rgb}{0.76,0.76,0.76}
\definecolor{gray77}{rgb}{0.77,0.77,0.77}
\definecolor{gray78}{rgb}{0.78,0.78,0.78}
\definecolor{gray79}{rgb}{0.79,0.79,0.79}
\definecolor{gray7}{rgb}{0.07,0.07,0.07}
\definecolor{gray80}{rgb}{0.80,0.80,0.80}
\definecolor{gray81}{rgb}{0.81,0.81,0.81}
\definecolor{gray82}{rgb}{0.82,0.82,0.82}
\definecolor{gray83}{rgb}{0.83,0.83,0.83}
\definecolor{gray84}{rgb}{0.84,0.84,0.84}
\definecolor{gray85}{rgb}{0.85,0.85,0.85}
\definecolor{gray86}{rgb}{0.86,0.86,0.86}
\definecolor{gray87}{rgb}{0.87,0.87,0.87}
\definecolor{gray88}{rgb}{0.88,0.88,0.88}
\definecolor{gray89}{rgb}{0.89,0.89,0.89}
\definecolor{gray8}{rgb}{0.08,0.08,0.08}
\definecolor{gray90}{rgb}{0.90,0.90,0.90}
\definecolor{gray91}{rgb}{0.91,0.91,0.91}
\definecolor{gray92}{rgb}{0.92,0.92,0.92}
\definecolor{gray93}{rgb}{0.93,0.93,0.93}
\definecolor{gray94}{rgb}{0.94,0.94,0.94}
\definecolor{gray95}{rgb}{0.95,0.95,0.95}
\definecolor{gray96}{rgb}{0.96,0.96,0.96}
\definecolor{gray97}{rgb}{0.97,0.97,0.97}
\definecolor{gray98}{rgb}{0.98,0.98,0.98}
\definecolor{gray99}{rgb}{0.99,0.99,0.99}
\definecolor{gray9}{rgb}{0.09,0.09,0.09}
\definecolor{gray}{rgb}{0.75,0.75,0.75}
\definecolor{green1}{rgb}{0.00,1.00,0.00}
\definecolor{green2}{rgb}{0.00,0.93,0.00}
\definecolor{green3}{rgb}{0.00,0.80,0.00}
\definecolor{green4}{rgb}{0.00,0.55,0.00}
\definecolor{greenyellow}{rgb}{0.68,1.00,0.18}
\definecolor{green}{rgb}{0.00,1.00,0.00}
\definecolor{grey0}{rgb}{0.00,0.00,0.00}
\definecolor{grey100}{rgb}{1.00,1.00,1.00}
\definecolor{grey10}{rgb}{0.10,0.10,0.10}
\definecolor{grey11}{rgb}{0.11,0.11,0.11}
\definecolor{grey12}{rgb}{0.12,0.12,0.12}
\definecolor{grey13}{rgb}{0.13,0.13,0.13}
\definecolor{grey14}{rgb}{0.14,0.14,0.14}
\definecolor{grey15}{rgb}{0.15,0.15,0.15}
\definecolor{grey16}{rgb}{0.16,0.16,0.16}
\definecolor{grey17}{rgb}{0.17,0.17,0.17}
\definecolor{grey18}{rgb}{0.18,0.18,0.18}
\definecolor{grey19}{rgb}{0.19,0.19,0.19}
\definecolor{grey1}{rgb}{0.01,0.01,0.01}
\definecolor{grey20}{rgb}{0.20,0.20,0.20}
\definecolor{grey21}{rgb}{0.21,0.21,0.21}
\definecolor{grey22}{rgb}{0.22,0.22,0.22}
\definecolor{grey23}{rgb}{0.23,0.23,0.23}
\definecolor{grey24}{rgb}{0.24,0.24,0.24}
\definecolor{grey25}{rgb}{0.25,0.25,0.25}
\definecolor{grey26}{rgb}{0.26,0.26,0.26}
\definecolor{grey27}{rgb}{0.27,0.27,0.27}
\definecolor{grey28}{rgb}{0.28,0.28,0.28}
\definecolor{grey29}{rgb}{0.29,0.29,0.29}
\definecolor{grey2}{rgb}{0.02,0.02,0.02}
\definecolor{grey30}{rgb}{0.30,0.30,0.30}
\definecolor{grey31}{rgb}{0.31,0.31,0.31}
\definecolor{grey32}{rgb}{0.32,0.32,0.32}
\definecolor{grey33}{rgb}{0.33,0.33,0.33}
\definecolor{grey34}{rgb}{0.34,0.34,0.34}
\definecolor{grey35}{rgb}{0.35,0.35,0.35}
\definecolor{grey36}{rgb}{0.36,0.36,0.36}
\definecolor{grey37}{rgb}{0.37,0.37,0.37}
\definecolor{grey38}{rgb}{0.38,0.38,0.38}
\definecolor{grey39}{rgb}{0.39,0.39,0.39}
\definecolor{grey3}{rgb}{0.03,0.03,0.03}
\definecolor{grey40}{rgb}{0.40,0.40,0.40}
\definecolor{grey41}{rgb}{0.41,0.41,0.41}
\definecolor{grey42}{rgb}{0.42,0.42,0.42}
\definecolor{grey43}{rgb}{0.43,0.43,0.43}
\definecolor{grey44}{rgb}{0.44,0.44,0.44}
\definecolor{grey45}{rgb}{0.45,0.45,0.45}
\definecolor{grey46}{rgb}{0.46,0.46,0.46}
\definecolor{grey47}{rgb}{0.47,0.47,0.47}
\definecolor{grey48}{rgb}{0.48,0.48,0.48}
\definecolor{grey49}{rgb}{0.49,0.49,0.49}
\definecolor{grey4}{rgb}{0.04,0.04,0.04}
\definecolor{grey50}{rgb}{0.50,0.50,0.50}
\definecolor{grey51}{rgb}{0.51,0.51,0.51}
\definecolor{grey52}{rgb}{0.52,0.52,0.52}
\definecolor{grey53}{rgb}{0.53,0.53,0.53}
\definecolor{grey54}{rgb}{0.54,0.54,0.54}
\definecolor{grey55}{rgb}{0.55,0.55,0.55}
\definecolor{grey56}{rgb}{0.56,0.56,0.56}
\definecolor{grey57}{rgb}{0.57,0.57,0.57}
\definecolor{grey58}{rgb}{0.58,0.58,0.58}
\definecolor{grey59}{rgb}{0.59,0.59,0.59}
\definecolor{grey5}{rgb}{0.05,0.05,0.05}
\definecolor{grey60}{rgb}{0.60,0.60,0.60}
\definecolor{grey61}{rgb}{0.61,0.61,0.61}
\definecolor{grey62}{rgb}{0.62,0.62,0.62}
\definecolor{grey63}{rgb}{0.63,0.63,0.63}
\definecolor{grey64}{rgb}{0.64,0.64,0.64}
\definecolor{grey65}{rgb}{0.65,0.65,0.65}
\definecolor{grey66}{rgb}{0.66,0.66,0.66}
\definecolor{grey67}{rgb}{0.67,0.67,0.67}
\definecolor{grey68}{rgb}{0.68,0.68,0.68}
\definecolor{grey69}{rgb}{0.69,0.69,0.69}
\definecolor{grey6}{rgb}{0.06,0.06,0.06}
\definecolor{grey70}{rgb}{0.70,0.70,0.70}
\definecolor{grey71}{rgb}{0.71,0.71,0.71}
\definecolor{grey72}{rgb}{0.72,0.72,0.72}
\definecolor{grey73}{rgb}{0.73,0.73,0.73}
\definecolor{grey74}{rgb}{0.74,0.74,0.74}
\definecolor{grey75}{rgb}{0.75,0.75,0.75}
\definecolor{grey76}{rgb}{0.76,0.76,0.76}
\definecolor{grey77}{rgb}{0.77,0.77,0.77}
\definecolor{grey78}{rgb}{0.78,0.78,0.78}
\definecolor{grey79}{rgb}{0.79,0.79,0.79}
\definecolor{grey7}{rgb}{0.07,0.07,0.07}
\definecolor{grey80}{rgb}{0.80,0.80,0.80}
\definecolor{grey81}{rgb}{0.81,0.81,0.81}
\definecolor{grey82}{rgb}{0.82,0.82,0.82}
\definecolor{grey83}{rgb}{0.83,0.83,0.83}
\definecolor{grey84}{rgb}{0.84,0.84,0.84}
\definecolor{grey85}{rgb}{0.85,0.85,0.85}
\definecolor{grey86}{rgb}{0.86,0.86,0.86}
\definecolor{grey87}{rgb}{0.87,0.87,0.87}
\definecolor{grey88}{rgb}{0.88,0.88,0.88}
\definecolor{grey89}{rgb}{0.89,0.89,0.89}
\definecolor{grey8}{rgb}{0.08,0.08,0.08}
\definecolor{grey90}{rgb}{0.90,0.90,0.90}
\definecolor{grey91}{rgb}{0.91,0.91,0.91}
\definecolor{grey92}{rgb}{0.92,0.92,0.92}
\definecolor{grey93}{rgb}{0.93,0.93,0.93}
\definecolor{grey94}{rgb}{0.94,0.94,0.94}
\definecolor{grey95}{rgb}{0.95,0.95,0.95}
\definecolor{grey96}{rgb}{0.96,0.96,0.96}
\definecolor{grey97}{rgb}{0.97,0.97,0.97}
\definecolor{grey98}{rgb}{0.98,0.98,0.98}
\definecolor{grey99}{rgb}{0.99,0.99,0.99}
\definecolor{grey9}{rgb}{0.09,0.09,0.09}
\definecolor{grey}{rgb}{0.75,0.75,0.75}
\definecolor{honeydew1}{rgb}{0.94,1.00,0.94}
\definecolor{honeydew2}{rgb}{0.88,0.93,0.88}
\definecolor{honeydew3}{rgb}{0.76,0.80,0.76}
\definecolor{honeydew4}{rgb}{0.51,0.55,0.51}
\definecolor{honeydew}{rgb}{0.94,1.00,0.94}
\definecolor{hotpink}{rgb}{1.00,0.41,0.71}
\definecolor{indianred}{rgb}{0.80,0.36,0.36}
\definecolor{ivory1}{rgb}{1.00,1.00,0.94}
\definecolor{ivory2}{rgb}{0.93,0.93,0.88}
\definecolor{ivory3}{rgb}{0.80,0.80,0.76}
\definecolor{ivory4}{rgb}{0.55,0.55,0.51}
\definecolor{ivory}{rgb}{1.00,1.00,0.94}
\definecolor{khaki1}{rgb}{1.00,0.96,0.56}
\definecolor{khaki2}{rgb}{0.93,0.90,0.52}
\definecolor{khaki3}{rgb}{0.80,0.78,0.45}
\definecolor{khaki4}{rgb}{0.55,0.53,0.31}
\definecolor{khaki}{rgb}{0.94,0.90,0.55}
\definecolor{lavenderblush}{rgb}{1.00,0.94,0.96}
\definecolor{lavender}{rgb}{0.90,0.90,0.98}
\definecolor{lawngreen}{rgb}{0.49,0.99,0.00}
\definecolor{lemonchiffon}{rgb}{1.00,0.98,0.80}
\definecolor{lightblue}{rgb}{0.68,0.85,0.90}
\definecolor{lightcoral}{rgb}{0.94,0.50,0.50}
\definecolor{lightcyan}{rgb}{0.88,1.00,1.00}
\definecolor{lightgoldenrod}{rgb}{0.93,0.87,0.51}
\definecolor{lightgoldenrod}{rgb}{0.98,0.98,0.82}
\definecolor{lightgray}{rgb}{0.83,0.83,0.83}
\definecolor{lightgreen}{rgb}{0.56,0.93,0.56}
\definecolor{lightgrey}{rgb}{0.83,0.83,0.83}
\definecolor{lightpink}{rgb}{1.00,0.71,0.76}
\definecolor{lightsalmon}{rgb}{1.00,0.63,0.48}
\definecolor{lightsea}{rgb}{0.13,0.70,0.67}
\definecolor{lightsky}{rgb}{0.53,0.81,0.98}
\definecolor{lightslate}{rgb}{0.47,0.53,0.60}
\definecolor{lightslate}{rgb}{0.47,0.53,0.60}
\definecolor{lightslate}{rgb}{0.52,0.44,1.00}
\definecolor{lightsteel}{rgb}{0.69,0.77,0.87}
\definecolor{lightyellow}{rgb}{1.00,1.00,0.88}
\definecolor{limegreen}{rgb}{0.20,0.80,0.20}
\definecolor{linen}{rgb}{0.98,0.94,0.90}
\definecolor{magenta1}{rgb}{1.00,0.00,1.00}
\definecolor{magenta2}{rgb}{0.93,0.00,0.93}
\definecolor{magenta3}{rgb}{0.80,0.00,0.80}
\definecolor{magenta4}{rgb}{0.55,0.00,0.55}
\definecolor{magenta}{rgb}{1.00,0.00,1.00}
\definecolor{maroon1}{rgb}{1.00,0.20,0.70}
\definecolor{maroon2}{rgb}{0.93,0.19,0.65}
\definecolor{maroon3}{rgb}{0.80,0.16,0.56}
\definecolor{maroon4}{rgb}{0.55,0.11,0.38}
\definecolor{maroon}{rgb}{0.69,0.19,0.38}
\definecolor{mediumaquamarine}{rgb}{0.40,0.80,0.67}
\definecolor{mediumblue}{rgb}{0.00,0.00,0.80}
\definecolor{mediumorchid}{rgb}{0.73,0.33,0.83}
\definecolor{mediumpurple}{rgb}{0.58,0.44,0.86}
\definecolor{mediumsea}{rgb}{0.24,0.70,0.44}
\definecolor{mediumslate}{rgb}{0.48,0.41,0.93}
\definecolor{mediumspring}{rgb}{0.00,0.98,0.60}
\definecolor{mediumturquoise}{rgb}{0.28,0.82,0.80}
\definecolor{mediumviolet}{rgb}{0.78,0.08,0.52}
\definecolor{midnightblue}{rgb}{0.10,0.10,0.44}
\definecolor{mintcream}{rgb}{0.96,1.00,0.98}
\definecolor{mistyrose}{rgb}{1.00,0.89,0.88}
\definecolor{moccasin}{rgb}{1.00,0.89,0.71}
\definecolor{navajowhite}{rgb}{1.00,0.87,0.68}
\definecolor{navyblue}{rgb}{0.00,0.00,0.50}
\definecolor{navy}{rgb}{0.00,0.00,0.50}
\definecolor{oldlace}{rgb}{0.99,0.96,0.90}
\definecolor{olivedrab}{rgb}{0.42,0.56,0.14}
\definecolor{orange1}{rgb}{1.00,0.65,0.00}
\definecolor{orange2}{rgb}{0.93,0.60,0.00}
\definecolor{orange3}{rgb}{0.80,0.52,0.00}
\definecolor{orange4}{rgb}{0.55,0.35,0.00}
\definecolor{orangered}{rgb}{1.00,0.27,0.00}
\definecolor{orange}{rgb}{1.00,0.65,0.00}
\definecolor{orchid1}{rgb}{1.00,0.51,0.98}
\definecolor{orchid2}{rgb}{0.93,0.48,0.91}
\definecolor{orchid3}{rgb}{0.80,0.41,0.79}
\definecolor{orchid4}{rgb}{0.55,0.28,0.54}
\definecolor{orchid}{rgb}{0.85,0.44,0.84}
\definecolor{palegoldenrod}{rgb}{0.93,0.91,0.67}
\definecolor{palegreen}{rgb}{0.60,0.98,0.60}
\definecolor{paleturquoise}{rgb}{0.69,0.93,0.93}
\definecolor{paleviolet}{rgb}{0.86,0.44,0.58}
\definecolor{papayawhip}{rgb}{1.00,0.94,0.84}
\definecolor{peachpuff}{rgb}{1.00,0.85,0.73}
\definecolor{peru}{rgb}{0.80,0.52,0.25}
\definecolor{pink1}{rgb}{1.00,0.71,0.77}
\definecolor{pink2}{rgb}{0.93,0.66,0.72}
\definecolor{pink3}{rgb}{0.80,0.57,0.62}
\definecolor{pink4}{rgb}{0.55,0.39,0.42}
\definecolor{pink}{rgb}{1.00,0.75,0.80}
\definecolor{plum1}{rgb}{1.00,0.73,1.00}
\definecolor{plum2}{rgb}{0.93,0.68,0.93}
\definecolor{plum3}{rgb}{0.80,0.59,0.80}
\definecolor{plum4}{rgb}{0.55,0.40,0.55}
\definecolor{plum}{rgb}{0.87,0.63,0.87}
\definecolor{powderblue}{rgb}{0.69,0.88,0.90}
\definecolor{purple1}{rgb}{0.61,0.19,1.00}
\definecolor{purple2}{rgb}{0.57,0.17,0.93}
\definecolor{purple3}{rgb}{0.49,0.15,0.80}
\definecolor{purple4}{rgb}{0.33,0.10,0.55}
\definecolor{purple}{rgb}{0.63,0.13,0.94}
\definecolor{red1}{rgb}{1.00,0.00,0.00}
\definecolor{red2}{rgb}{0.93,0.00,0.00}
\definecolor{red3}{rgb}{0.80,0.00,0.00}
\definecolor{red4}{rgb}{0.55,0.00,0.00}
\definecolor{red}{rgb}{1.00,0.00,0.00}
\definecolor{rosybrown}{rgb}{0.74,0.56,0.56}
\definecolor{royalblue}{rgb}{0.25,0.41,0.88}
\definecolor{saddlebrown}{rgb}{0.55,0.27,0.07}
\definecolor{salmon1}{rgb}{1.00,0.55,0.41}
\definecolor{salmon2}{rgb}{0.93,0.51,0.38}
\definecolor{salmon3}{rgb}{0.80,0.44,0.33}
\definecolor{salmon4}{rgb}{0.55,0.30,0.22}
\definecolor{salmon}{rgb}{0.98,0.50,0.45}
\definecolor{sandybrown}{rgb}{0.96,0.64,0.38}
\definecolor{seagreen}{rgb}{0.18,0.55,0.34}
\definecolor{seashell1}{rgb}{1.00,0.96,0.93}
\definecolor{seashell2}{rgb}{0.93,0.90,0.87}
\definecolor{seashell3}{rgb}{0.80,0.77,0.75}
\definecolor{seashell4}{rgb}{0.55,0.53,0.51}
\definecolor{seashell}{rgb}{1.00,0.96,0.93}
\definecolor{sienna1}{rgb}{1.00,0.51,0.28}
\definecolor{sienna2}{rgb}{0.93,0.47,0.26}
\definecolor{sienna3}{rgb}{0.80,0.41,0.22}
\definecolor{sienna4}{rgb}{0.55,0.28,0.15}
\definecolor{sienna}{rgb}{0.63,0.32,0.18}
\definecolor{skyblue}{rgb}{0.53,0.81,0.92}
\definecolor{slateblue}{rgb}{0.42,0.35,0.80}
\definecolor{slategray}{rgb}{0.44,0.50,0.56}
\definecolor{slategrey}{rgb}{0.44,0.50,0.56}
\definecolor{snow1}{rgb}{1.00,0.98,0.98}
\definecolor{snow2}{rgb}{0.93,0.91,0.91}
\definecolor{snow3}{rgb}{0.80,0.79,0.79}
\definecolor{snow4}{rgb}{0.55,0.54,0.54}
\definecolor{snow}{rgb}{1.00,0.98,0.98}
\definecolor{springgreen}{rgb}{0.00,1.00,0.50}
\definecolor{steelblue}{rgb}{0.27,0.51,0.71}
\definecolor{tan1}{rgb}{1.00,0.65,0.31}
\definecolor{tan2}{rgb}{0.93,0.60,0.29}
\definecolor{tan3}{rgb}{0.80,0.52,0.25}
\definecolor{tan4}{rgb}{0.55,0.35,0.17}
\definecolor{tan}{rgb}{0.82,0.71,0.55}
\definecolor{thistle1}{rgb}{1.00,0.88,1.00}
\definecolor{thistle2}{rgb}{0.93,0.82,0.93}
\definecolor{thistle3}{rgb}{0.80,0.71,0.80}
\definecolor{thistle4}{rgb}{0.55,0.48,0.55}
\definecolor{thistle}{rgb}{0.85,0.75,0.85}
\definecolor{tomato1}{rgb}{1.00,0.39,0.28}
\definecolor{tomato2}{rgb}{0.93,0.36,0.26}
\definecolor{tomato3}{rgb}{0.80,0.31,0.22}
\definecolor{tomato4}{rgb}{0.55,0.21,0.15}
\definecolor{tomato}{rgb}{1.00,0.39,0.28}
\definecolor{turquoise1}{rgb}{0.00,0.96,1.00}
\definecolor{turquoise2}{rgb}{0.00,0.90,0.93}
\definecolor{turquoise3}{rgb}{0.00,0.77,0.80}
\definecolor{turquoise4}{rgb}{0.00,0.53,0.55}
\definecolor{turquoise}{rgb}{0.25,0.88,0.82}
\definecolor{violetred}{rgb}{0.82,0.13,0.56}
\definecolor{violet}{rgb}{0.93,0.51,0.93}
\definecolor{wheat1}{rgb}{1.00,0.91,0.73}
\definecolor{wheat2}{rgb}{0.93,0.85,0.68}
\definecolor{wheat3}{rgb}{0.80,0.73,0.59}
\definecolor{wheat4}{rgb}{0.55,0.49,0.40}
\definecolor{wheat}{rgb}{0.96,0.87,0.70}
\definecolor{whitesmoke}{rgb}{0.96,0.96,0.96}
\definecolor{white}{rgb}{1.00,1.00,1.00}
\definecolor{yellow1}{rgb}{1.00,1.00,0.00}
\definecolor{yellow2}{rgb}{0.93,0.93,0.00}
\definecolor{yellow3}{rgb}{0.80,0.80,0.00}
\definecolor{yellow4}{rgb}{0.55,0.55,0.00}
\definecolor{yellowgreen}{rgb}{0.60,0.80,0.20}
\definecolor{yellow}{rgb}{1.00,1.00,0.00}

\usepackage{xcolor}
\usepackage{hyperref}
\definecolor{DarkGreen}{rgb}{0.00,0.39,0.00}

\usepackage{amsmath,amsfonts}
\usepackage{amssymb} 
\shorttitle{Rearrangement of secondary flow over spanwise heterogeneous roughness}
\shortauthor{A. Stroh, K. Sch\"{a}fer, B. Frohnapfel and P. Forooghi}

\title{Rearrangement of secondary flow over spanwise heterogeneous roughness}

\author{A. Stroh 
  \corresp{\email{alexander.stroh@kit.edu}},
  K. Sch\"{a}fer, 
  B. Frohnapfel 
 \and P. Forooghi 
 }

\affiliation{ 
Institute of Fluid Mechanics, Karlsruhe Institute of Technology, Karlsruhe, Germany}

\begin{document}

\maketitle

\begin{abstract}
Turbulent flow over a surface with streamwise-elongated rough and smooth stripes is studied by means of direct numerical simulation (DNS) in a periodic \textcolor{black}{plane open channel} with fully resolved roughness. 
The goal is to understand how the mean height of roughness affects the characteristics of the secondary flow formed above a spanwise-heterogeneous rough surface. 
To this end, while the \textcolor{black}{statistical properties of roughness texture} as well as the width and spacing of the rough stripes are kept constant, the elevation of the smooth stripes is systematically varied in different simulation cases. 
Utilizing this variation three configurations representing protruding, recessed and an intermediate type of roughness are analysed. In all cases secondary flows are present and the  skin friction coefficients calculated for all the heterogeneous rough surfaces are meaningfully larger than what would result from the area-weighted average of those of homogeneous smooth and rough surfaces. This drag increase appears to be linked to the strength of the secondary flow.
The rotational direction of the secondary motion is shown to depend on the relative surface elevation. The present results suggest that this rearrangement of the secondary flow is linked to the spatial distribution of the spanwise-wall-normal Reynolds stress component which  carries opposing signs for protruding and recessed roughness.

\end{abstract}


\section{Introduction}

Occurrence of a pronounced fluid motion perpendicular to the main flow direction has been observed in various wall-bounded flow configurations. 
 \cite{prandtl1931einfuhrung} introduced the term \textit{secondary flows} for this phenomenon and categorized them into three kinds.
Secondary motions of \textit{Prandtl's second kind}, which are in the focus of the present paper, occur in turbulent flows and are related to inhomogeneities of the Reynolds stresses. The classical example of this kind of secondary motion is flow in ducts with non-circular cross sections, which was first reported by \cite{nikuradse1926untersuchung}.
In spite of its weak intensity, the secondary motion is  known to be able to noticeably deform the primary mean velocity profile.
Secondary flow of {Prandtl's second kind} can also occur in plane or symmetrical wall-bounded flows (i.e. channels or pipes) 
if a local spanwise inhomogeneity in wall conditions is present, due to \textit{e.g.} surface roughness. 
The pioneering work by \cite{Hinze1967,Hinze1973} demonstrated the formation of secondary motions over flow-aligned roughness stripes in a duct with an upwelling motion above the smooth wall and a downwelling motion over the rough parts. 

This phenomenon is particularly relevant in applications with spatially non-uniform roughness formation, a well-documented example being flow over turbomachinery blades \citep{Bons2001}. Experimental investigation of turbulent flow over a damaged turbine blade with irregular surface roughness by \citet{Mejia2013} and \citet{Barros2014} clearly showed the formation of secondary motions over such a blade surface.
Another important manifestation of these secondary motions occurs in river flows, where lateral sediment transport can reinforce and maintain spanwise surface variations of longitudinal bedforms \citep{Wang2006}. 


Despite the great variety of possible configurations with lateral wall inhomogeneity, two main configurations 
have been well studied in the recent decades: a spanwise variation in wall shear stress and a spanwise variation of local elevation of the wall. 
Following \citet{Wang2006}, we refer to the configuration where the former effect is dominant 
as \textit{strip-type roughness} and to the latter as \textit{ridge-type roughness}, respectively.
These authors observed upwelling and downwelling motion in strip-type roughness to occur above the smooth and rough stripes, while for ridge-type roughness, the upwelling and downwelling motions were observed above the elevated and recessed wall areas, respectively.

\citet{Willingham2014} and \citet{Chung2018similarity} numerically studied idealised strip-type roughness in plane channels with stripes of low and high imposed friction drag on the wall surface.
Both groups report a similar secondary flow to that observed by \cite{Hinze1967,Hinze1973}, \textit{i.e.} upwelling motion over the low-shear and downwelling motion over the high-shear region. 
{\textcolor{black} {This behavior was linked to the experimental observations for flows over 
 damaged turbine blades for which \citet{Mejia2013} and \citet{Barros2014} identified regions with low and high mean streamwise velocity, termed as low and  high momentum pathways (LMP and HMP), which are flanked by streamwise-oriented swirling motions \citep{Anderson2015}.}}
In the study of \citet{Chung2018similarity} it is shown that either LMP or HMP can be located above the high-shear stripe depending on the spanwise \textcolor{black}{extent} of the stripe. For stripes of free-slip and no-slip boundary conditions \citet{Tuerk2014} and \citet{Stroh2016} reported a switch of the secondary motion rotational sense through a variation of the spanwise \textcolor{black}{extent} of the free-slip region.
Ridge-type roughness has also been studied by various research groups.
\citet{Goldstein1998secondary} investigated secondary flow above riblets concluding that it is mainly caused by the deflection of the spanwise velocity fluctuations.
\cite{Vanderwel2015} and \cite{Vanderwel2019} studied turbulent flow over streamwise elongated rows of Lego blocks, both experimentally and numerically. 
\cite{hwang2018secondary} employed DNS to examine turbulent boundary layer over streamwise aligned ridges.
All these groups observed an upwelling motion above the protruding surface areas.

Despite the fact that reduction of the possible structure configuration to strip- and ridge-type roughness facilitates physical understanding by isolating the shear-increasing effect of the surface roughness from the effects linked to wall elevation, 
one should note that a clear-cut separation between the two categories is not necessarily possible for realistic roughness. 
The reason is that formation of roughness  is inherently accompanied by a change in the surface height. 
This calls for an understanding of the conditions under which the behavior of the secondary flow over a realistic roughness resembles that of each category. 
Such an open question deserves particular attention in case of a protruding roughness, where the shear-inducing and height-increasing effects of roughness can cause opposing senses of rotation in the idealized scenarios. 

The aim of the present work is to systematically investigate the effect of roughness mean height on secondary motions and thereby provide an understanding on the conditions under which a roughness stripe can be classified under ridge- or strip-type roughness. 
The effect of spanwise spacing 
is not part of this investigation.
To this end we use DNS to study turbulent flow in \textcolor{black}{an open } channel with streamwise elongated stripes of roughness, and at the same time systematically vary the mean height of roughness related to the smooth wall level. 
\textcolor{black}{The roughness topography is nearly identical in statistical sense in all cases and the height difference is varied by shifting the smooth wall.}

\section{Procedure}

\begin{figure}
\centering
\includegraphics{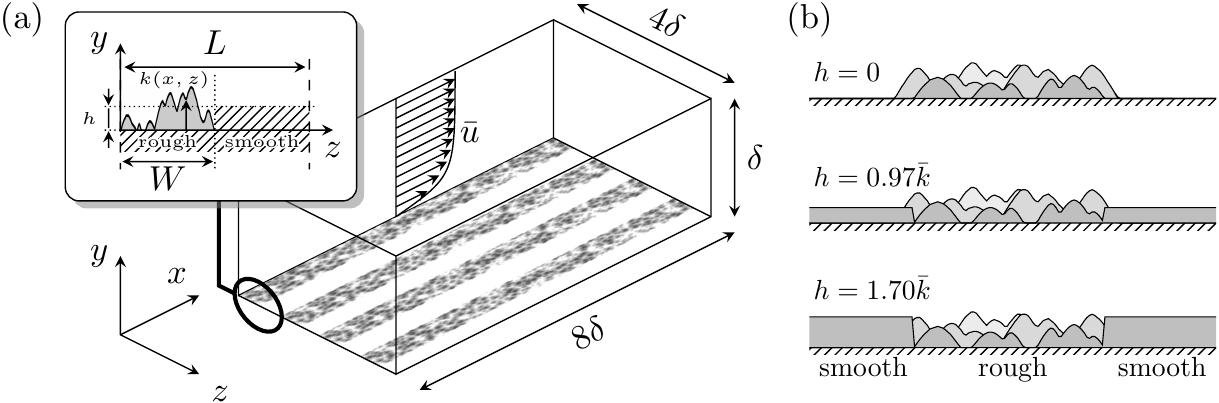}
\caption{Schematic of the \textcolor{black}{open channel} numerical domain with roughness stripes at the walls (a) and  introduced variation of the smooth wall elevation (b).\label{fig:numscheme}}
\end{figure}


A series of DNS has been carried out in a fully developed turbulent open channel flow driven by constant pressure gradient (CPG).
\textcolor{black}{
The Navier-Stokes equations are numerically integrated using the spectral solver SIMSON~\citep{simson2} which employs Fourier decomposition in the horizontal directions and Chebyshev discretization in the wall-normal~direction.}
A schematic of the numerical domain is depicted in Figure~\ref{fig:numscheme}(a). 
Periodic boundary conditions are applied in streamwise ($x$) and spanwise directions ($z$), while the wall-normal extension of the domain ($y$) is bounded by no-slip boundary conditions at the lower domain wall ($y = 0$) and symmetry boundary condition ($v = 0$, $\partial u/\partial y = \partial w / \partial y = 0$) at the upper boundary ($y = \delta$).
The numerical domain with the size of $(L_x \times L_y \times L_z) = (8 \delta \times \delta \times 4\delta)$ is discretized with {$768 \times 301 \times 384$} grid nodes resulting in a spatial resolution of $(\Delta x^+ \times \Delta y_{min}^+, \Delta y_{max}^+ \times \Delta z^+) = (5.2 \times 0.014, 2.6 \times 5.2)$.
The velocity components in the streamwise, wall-normal and spanwise direction are denoted by $(u,v,w)$, respectively.

Statistical integration in time and streamwise direction is carried out over approximately 50 flow-through times for every considered simulation configuration. 
The initial transient after the introduction of a structured surface is excluded from this statistical integration,
\textcolor{black}{i.e. the integration is started approximately two flow-through times  after the bulk mean velocity reached a statistically steady state.} 
The decomposition of the velocity field into mean part and
fluctuations given as $u_i (x,y,z,t) = \bar{u}_i (y,z) + u^{\prime}_i (x,y,z,t)$ is utilized. 
Hereby the quantities averaged in streamwise direction and time are denoted with an overbar $\bar{(\cdot)}$, while angular brackets $\left< \cdot \right>$ denote averaging in spanwise direction. 
Additionally, based on the assumption of a symmetric velocity distribution with respect to the middle of the rough or the smooth surface stripes, we use those symmetries in the averaging procedure in order to obtain smoother statistical data.

\begin{figure}
    \centering
\includegraphics[]{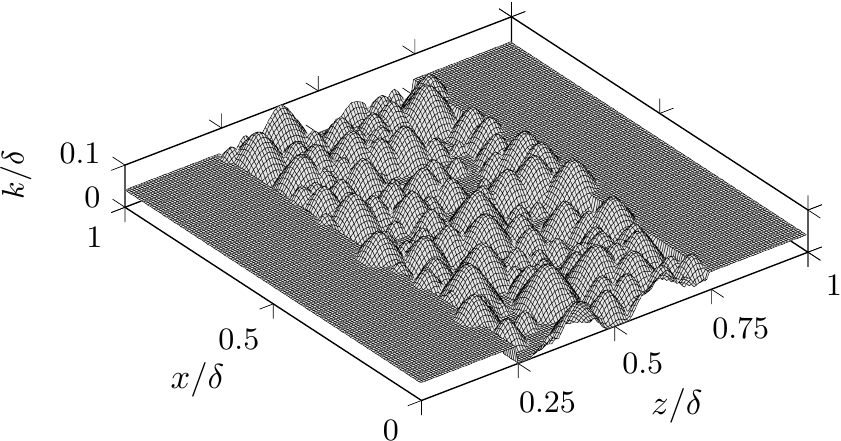}
    \caption{Zoomed view on the three-dimensional roughness distribution at $h=0.97\bar{k}$.}
    \label{fig:3d}
\end{figure}

The rough and elevated smooth surfaces are modeled by introduction of an external volume force field to
the Navier-Stokes equations based on the immersed boundary method proposed by \citet{Goldstein_1993}.
The presently used immersed boundary implementation has been validated in previous studies by \cite{forooghi2018direct} and \cite{Vanderwel2019}.
The wave-length, $L$, represents the size of the alternating structure with a constant roughness fraction $\Phi=W/L=0.5$, where $W$ denotes the width of the rough area (Figure~\ref{fig:numscheme},a).
Based on literature results \textcolor{black}{\citep{Vanderwel2015, hwang2018secondary,Chung2018similarity} }
the wave-length $L/\delta = 1$ is considered, for which the formation of a strong large-scale secondary motion with pronounced LMPs \& HMPs is expected.
\textcolor{black}{The rough surface is generated using the technique proposed by \cite{Forooghi2017}, in which several discrete roughness elements are distributed randomly on the bottom surface, creating a rough surface with certain statistics. In the present simulations all roughness stripes have virtually the same statistical properties. Considering a homogeneous roughness the statistical properties are as follow; mean elevation $\bar{k}/\delta=0.043$, maximum peak elevation \mbox{$k_\text{max}/\delta=0.10$}, root mean square elevation $k_\text{rms}/\delta=0.024$, skewness $Sk=0.079$ and kurtosis $Ku=2.24$. One should note that these statistical properties belong to the rough areas and not the entire surface. Figure~\ref{fig:3d}  shows a zoomed view of a roughness stripe for one of the cases ($h=0.97\bar{k}$). As can be seen in the figure, there is a gradual transition from the smooth to the rough region. For this purpose, initially the elements are  distributed on a wider area than the intended stripe width $W$, and consequently, all the elements whose centres lie beyond the intended border are eliminated.} 
As an additional reference case we also carry out a simulation, where the entire wall area is uniformly covered by the rough surface.
\textcolor{black}{The border treatment at the edge of rough stripes slightly modifies the statistical properties of the roughness compared to the uniformly rough surface.}
Figure~\ref{fig:numscheme}(b) shows the three considered elevations of the smooth surface: $h=0, 0.97\bar{k}$ and $1.70\bar{k}$.
These three values might represent different roughness types \citep{Bons2001}: roughness generated by deposition ($h=0$, positively skewed, protruding roughness), roughness generated by simultaneous deposition and erosion ($h=0.97\bar{k}$, near zero skewness) and roughness generated by pitting, erosion or corrosion ($h=1.70\bar{k}$, negatively skewed, recessed or "carved" roughness).

\textcolor{black}{
The friction Reynolds number in all simulations is fixed at {$\mathrm{Re}_\tau= \delta_\text{eff} / \delta_\nu= 500$} with
the viscous lengthscale $\delta_\nu = \nu / u_\tau$,
}
the friction velocity $u_\tau=\sqrt{{\tau_\text{eff}}/{\rho}}$ and the effective wall-shear stress $\tau_\text{eff}=-\delta_\text{eff} P_x$.
The effective channel half-height $\delta_\text{eff} = \delta - h_\text{eff}$ (shown in Table~\ref{table:intprop} for every configuration) takes into account the reduction of the cross-sectional area of the channel, where $h_\text{eff}$ denotes the melt-down height of the entire introduced surface structure 
and $P_x$ is the imposed streamwise pressure gradient.
Due to the reduction of effective channel half-height for structured channels  ($\delta_\text{eff} < \delta$) the identical friction Reynolds number (and hence the same scale separation) is maintained across all simulations by 
\textcolor{black}{a reduction of $\delta_\nu$}
realized through an adjustment of the pressure gradient, so that $P_{x}=P_x^\text{s} \left( \delta / \delta_\text{eff} \right)^3$, where $P_x^\text{s}$ corresponds to the pressure gradient of the reference smooth channel simulation at $\mathrm{Re}_\tau=500$.
Since $\mathrm{Re}_\tau$ is fixed, the introduction of the structured surface into the flow field translates into a reduction of the bulk mean velocity 
$U_b = 1/(\delta_\text{eff} L_z) \int_0^{L_z} \int_0^\delta \bar{u} (y,z) \mathrm{d} y~\mathrm{d} z$ 
and  the corresponding bulk Reynolds number $\mathrm{Re}_b = {U_b \delta_\text{eff}}/{\nu}$.

Throughout the manuscript the nondimensionalization in viscous units is indicated by the superscript plus sign $(\cdot)^+$. It is performed using the friction velocity $u_\tau$ of the particular simulation. 
The superscript letters "s" and "r" denote the quantities of the smooth and homogeneous rough channel simulation, respectively.
\textcolor{black}{Extrinsic spatial averaging is utilized in the presented statistical datasets, i.e. the solid regions (with zero velocity) are included into the averaging procedure.}



\section{Results}

\subsection{Global Flow Properties}

\begin{table}
  \begin{center}
\def~{\hphantom{0}}
  \begin{tabular}{@{}ccccccccccc}
case & $\delta_\text{eff}/\delta$ &{$\mathrm{Re}_\tau$}& {$\mathrm{Re}_b$} & $U_b / U_b^s$ & {$U_b^+$}  & {$c_f/c_f^s$} & \tiny{$\left(\frac{\sqrt{\bar{v}^2+\bar{w}^2}}{U_b} \right)_\text{max}$}     
    & {$\Omega_x \delta_\text{eff}^2 /U_b^2$} \\[3pt]
	smooth & 1.000 & 500.0 & 9047 & 1.000  & 18.1  & $1.00$ & - &- \\
	$h=0$ & 0.978 & 499.6 & 5756 & 0.636 & 11.5 &  $2.47$ & $2.89 \cdot 10^{-2}$  &$1.46 \cdot 10^{-3}$  \\
	$h=0.97\bar{k}$ & 0.958 & 500.0 & 5981 & 0.661 & 12.0  & $2.29$ & $2.09 \cdot 10^{-2}$ & $0.57 \cdot 10^{-3}$ \\
	$h=1.70 \bar{k}$ & 0.943 & 500.0 & 6050 & 0.669 & 12.1  & $2.24$ & $3.03 \cdot 10^{-2}$ & $1.00 \cdot 10^{-3}$ \\
	rough & 0.957 & 500.4 & 5241 & 0.579 & 10.5 & $2.99$ & $1.13 \cdot 10^{-2}$ &$0.17 \cdot 10^{-3}$ \\
  \end{tabular}
  \caption{ Global flow properties for the considered configurations. \label{table:intprop}
  }
  \end{center}
\end{table}

Table~\ref{table:intprop} presents the global flow properties of the three considered heterogeneously rough configurations and compares them to a smooth and to a homogeneous rough channel flow. 
The spanwise heterogeneous rough surfaces, in which half of the total surface is covered by roughness,  exhibit a pronounced reduction of $U_b$ by $32-35\%$ in respect to the smooth case, while the homogeneous rough surface yields a reduction of $40\%$.
The observed augmentation in skin friction coefficient,
$c_f = 2 u_\tau^2 /  U_b^2$,
 primarily originates from this reduction in bulk velocity.

If drag on the heterogeneous roughness could be calculated by superposition of the smooth and an entirely rough surface, which is arguably the asymptotic case when the stripes are extremely wide, the result would be 
$c_f=0.5(c_f^s + c_f^r)=2.00c_f^s$.
In comparison to this asymptotic state $24\%, 18\%$ and $26\%$ higher $c_f$ is observed for $h=0, 0.97\bar{k}$ and $1.70\bar{k}$, respectively, indicating a significant impact of secondary motions on skin friction drag as previously discussed by e.g. \citet{Tuerk2014} or \citet{Chung2018similarity}.

The strength of the secondary motion can be measured in terms of the maximal magnitude of the induced secondary motion $\left( \sqrt{\bar{v}^2+\bar{w}^2} / U_b \right)_\text{max}$ or the 
specific mean streamwise enstrophy 
$\Omega_x = 1/A \int_0^{L_z} \int_{0}^\delta \bar{\omega}_x^2 \mathrm{d}y\mathrm{d}z$,
 where $\omega_x$ is the streamwise vorticity and $A$ is the cross-sectional area of the flow field, i.e. \textcolor{black}{$A = L_z \delta_\text{eff} = 4 \delta \delta_\text{eff}$}.
The latter can be understood as a measure of the rotational energy contained in the secondary motions. \citet{Stroh2016} showed that a minimum of the specific enstrophy can be linked to the reversal of secondary flow direction in case of stripes with slip and no-slip boundary conditions. This is also the case for the present data as will be discussed later. 
The weakest secondary motion in the present work is observed for $h= 0.97\bar{k} $ which is also the case with the weakest drag increase. For the other two cases with similarly large drag increase, the secondary motion magnitude appears to be a better qualitative measure for the impact of the secondary flow on skin friction drag. 
\textcolor{black}{It has to be noted that the homogeneous rough case also contains mean streamwise rotational energy, which is linked to the presence of local small-scale cross-sectional flows induced by roughness elements. 
The rotational energy content is, however, significantly smaller than the weakest secondary motion at $h=0.97\bar{k}$ and its appearance is limited to the near-wall region slightly extending beyond  $y=k_{max}$. 
}



\begin{figure}
\centering
\includegraphics[]{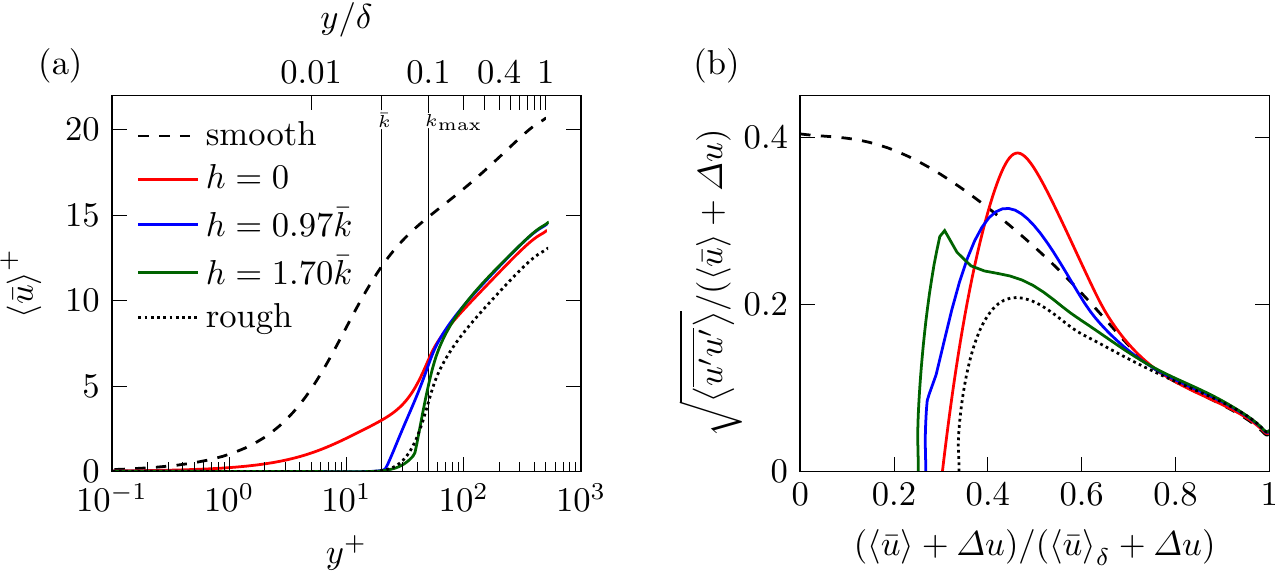}
\caption{Mean velocity profile in inner scaling in logarithmic form (a) 
and diagnostic plot scaled with $\left< \bar{u} \right> + \Delta u$ (b).
\label{fig:umean}}
\end{figure}

Figure~\ref{fig:umean}(a) presents the inner-scaled mean velocity profiles in logarithmic form for the considered simulation configurations.
The significant reduction of $U_b$ and the corresponding downward shift of the logarithmic region of the profile is evident for all rough cases. 
\textcolor{black}{
In order to check whether  these spanwise-averaged velocity profiles comply with outer layer similarity we consider the diagnostic plot as introduced by \citet{alfredsson2010diagnostic} in the adapted version for rough surfaces as proposed by \citet{castro2013outer}. To this end the roughness function  $\Delta u$ is extracted from  Figure~\ref{fig:umean}(a) and introduced in the normalization of the diagnostic function as shown in  Figure~\ref{fig:umean}(b). In this representation the effects of absolute wall distance and wall-shear stress are excluded such that the dynamic similarity of turbulence intensity and mean velocity can directly be compared among all cases. 
It can be observed that the streamwise velocity fluctuations linearly scale with the local mean streamwise velocity in all considered configurations for $0.7< (\left< \bar{u}\right> + \Delta u) /  (\left< \bar{u}\right>_\delta + \Delta u) < 0.9$.
In addition, all profiles  collapse onto the smooth wall case in the outer region thus indicating outer layer similarity of the different turbulent flows.
As previously reported by \citet{medjnoun2018characteristics}, this suggests that the observed secondary motion alters the spanwise averaged mean velocity profile and related turbulent fluctuations in the outer region in a similar manner for all considered simulations independent of the roughness properties. }

\subsection{Secondary Motion}

\begin{figure}
\centering
\includegraphics{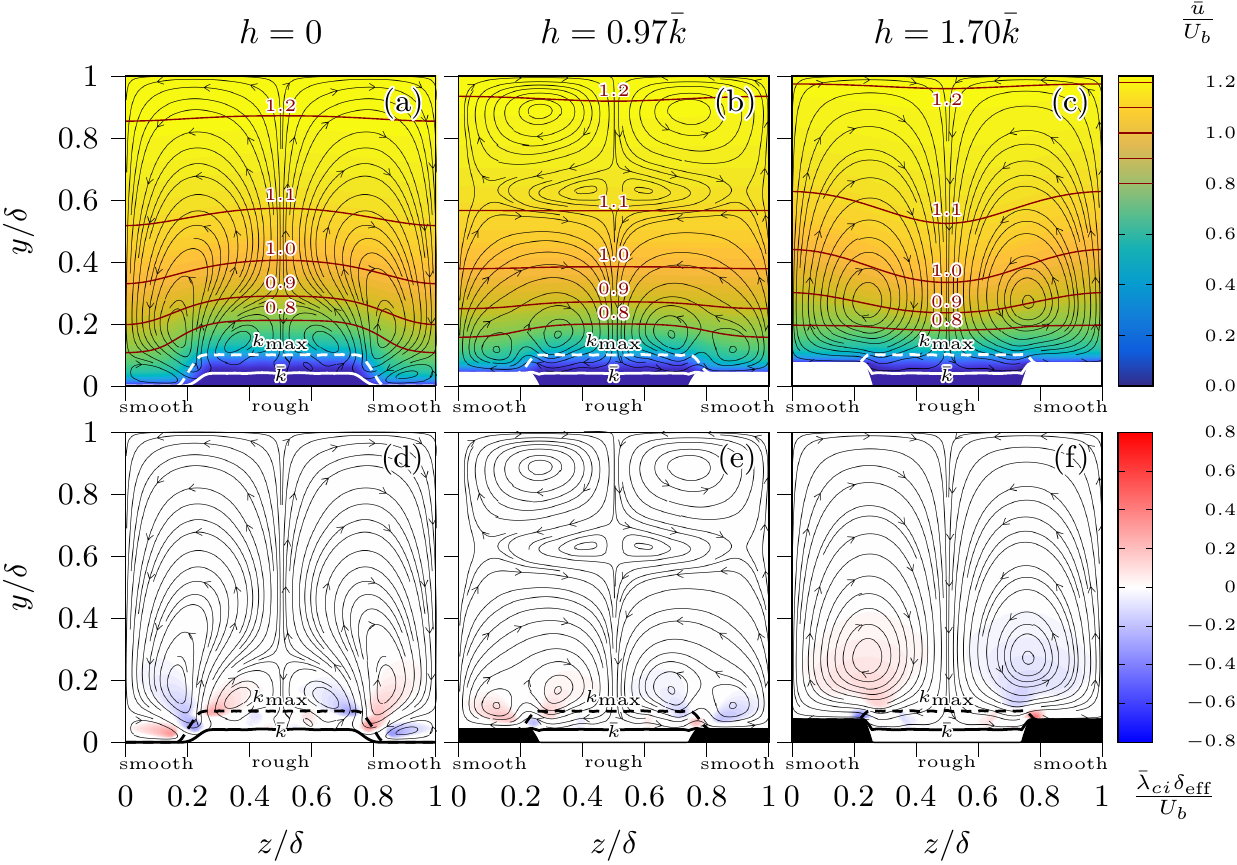}
\caption{Mean velocity profile  (a-c)
and sighned swirling strength (d-e) 
at different elevation of the smooth stripes $h$. Black lines indicate time-averaged streamlines of secondary motion in $y$-$z$-plane, brown solid lines mark the isolines of the streamwise mean velocity distribution. \label{fig:mean}}
\end{figure}

Figure~\ref{fig:mean} presents the distribution of the mean velocity overlayed with the secondary flow (depicted by in-plane streamlines) for the three considered elevations of the smooth wall.
In all three cases pronounced secondary motion patterns can be observed.
In the case of $h=0$ (Figure~\ref{fig:mean},a) the two main large-scale vortices  originate from the edges of the rough ridge. Two additional counter-rotating small vortex-pairs are located on the smooth wall and on top of the rough patch. The deformation of the streamwise velocity profile is shown with brown velocity isolines. It can be seen that a LMP is present over the rough surface part. 
This flow topology is similar to the secondary flows over ridge-type roughness (e.g. \citet{hwang2018secondary}).

In contrast, the case with  $h=1.70\bar{k}$ (Figure~\ref{fig:mean},c) shows a  downward bulging of the streamwise velocity field and thus HMP over the rough surface part. In this case the secondary motion is given through a single counter-rotating vortex pair with an upward motion above the elevated smooth region.  
This flow topology resembles the secondary flow reported for strip-type roughness (e.g. \citet{Willingham2014,Chung2018similarity}). 

The comparison of these two cases suggests that the alteration of the smooth wall elevation is an additional parameter for the secondary motion formation, which might enable rearrangement of the secondary flow topology from the ridge-type regime \textcolor{black}{(LMP over rough area)} to the strip-type regime (HMP over rough area).

The third case  $h=0.97\bar{k}$ (Figure~\ref{fig:mean},b) corresponds to  an 
intermediate state between ridge- and strip-type roughness. 
In this flow a more complex secondary flow topology is present. 
The largest vortical structures do not cover the entire vertical domain and are significantly weaker as indicated by the values listed in Table~\ref{table:intprop}.  
The streamwise enstrophy as well as the maximum magnitude of the secondary motion are lowest for this case. 
The rotational direction of the vortex pair in the lower channel half corresponds to the one observed for $h=1.70\bar{k}$ and the small one located in the center of the roughness for $h=0$. 

Note that even though the bulging pattern of the mean streamwise velocity contours is different in the three cases, it is not so strong as to disturb the similarity of the velocity defect profiles depicted in Figure~\ref{fig:umean}(b-d). This finding is in agreement with the suggestion by \cite{Chung2018similarity} that a departure from the global outer layer similarity -- or laterally uniform regime as referred to by these authors -- occurs when the ratio of roughness spacing to channel half-height, $W/\delta$, exceeds a threshold that is between 0.39 and 0.79; the present value is 0.5.

\subsection{Turbulent flow properties}
For typical rough surfaces shear stress and turbulent kinetic energy (TKE) in the vicinity of roughness is higher compared to a smooth wall at the same flow rate. In case of laterally heterogeneous roughness, spanwise gradients of these quantities are typically related to the occurrence of secondary motions \citep{Barros2014} even though it is still an open issue in literature whether HMP or LMP are located over high shear stress regions \citep{Chung2018similarity}. 

\begin{figure}
\centering
\includegraphics{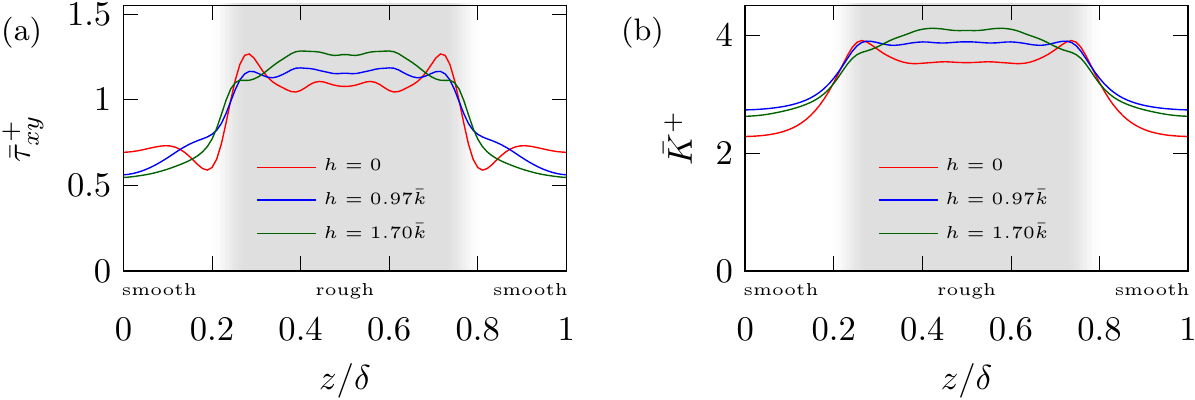}
\caption{Total stress (a) and turbulent kinetic energy (b) extracted at $y=k_\text{max}=0.1 \delta$ for the three different elevations of the smooth stripes. 
\label{fig:1dcomp}
}
\end{figure}

Figure~\ref{fig:1dcomp} shows the spanwise variation of total shear stress,  $\bar{\tau}_{xy}=\mu \frac{d\bar{u}}{dy}-\rho \overline{u^\prime v^\prime}$, and turbulent kinetic energy, $\bar{K}=0.5 (\overline{u^\prime u^\prime}+\overline{v^\prime v^\prime}+\overline{w^\prime w^\prime})$,
for all three cases of the present investigation at the same wall-normal location.
It can be seen that  all roughness stripes yield a similar distribution of total shear stress and TKE  in the sense that the regions with high levels of these quantities are always located above the rough stripes. 
Thus these flows are indeed examples where either HMP or LMP can be located above the high shear stress region depending on the relative height of the roughness.  
It can be deduced that spatial gradients of TKE or shear stress are not directly linked to the rotational direction of the largest secondary motion and the related occurrence of HMP and LMP in these cases. 
At the same time it should be noted that the secondary motions directly above the rough stripes are similar for all three cases, as they all generate a spanwise mean flow from the middle of the roughness patch towards its edges (\textcolor{black}{see Figure~\ref{fig:mean}}), which agrees with the rotational direction reported for strip-type roughness. 
This secondary motion appears to be strengthened further in case of $h=1.70\bar{k}$ while an opposing secondary motion, originating from the edges of the rough patch, dominates the case $h=0$.

Regarding the edge of protruding surface structures, \citet{hwang2018secondary} identified the wall-normal deflection of spanwise velocity fluctuations at this location -- which results in a strong correlation of spanwise and wall-normal velocity fluctuations, i.e. $\overline{v^\prime w^\prime}$ --  as an important quantity for the formation of secondary motions. 
The $\overline{v^\prime w^\prime}$ Reynolds stress and in particular its spatial gradients were also found to be important for the rearrangement of secondary flows over slip/no-slip stripes with varying width  \citep{Stroh2016}. 

\begin{figure}
\centering
\includegraphics{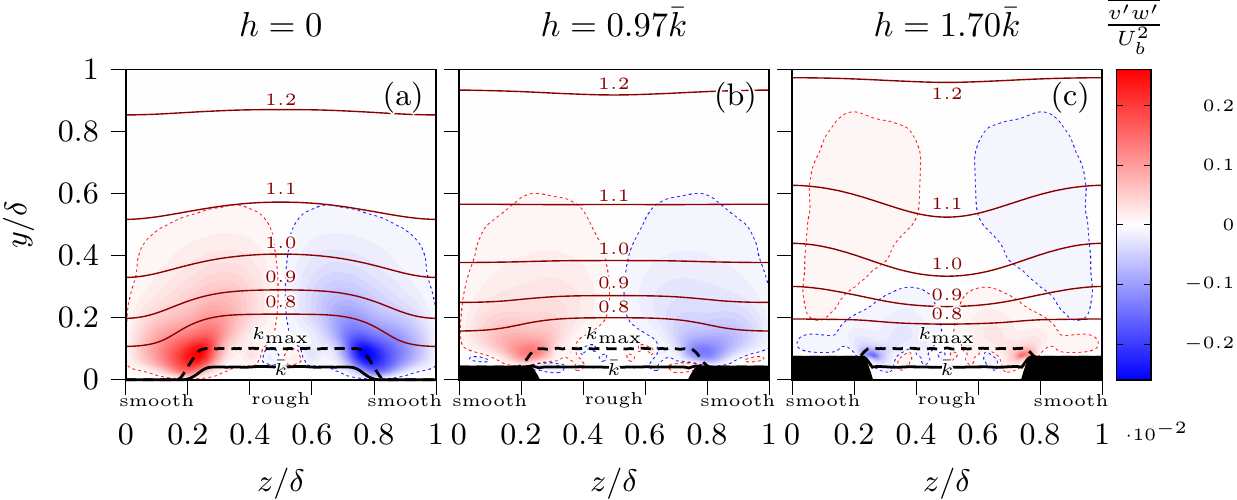}
\caption{Reynolds stress $\overline{v^\prime w^\prime}$ at different elevation of the smooth stripes $h$. Brown solid lines mark the isolines for the streamwise mean velocity distribution with isolevels corresponding to Figure~\ref{fig:mean}. \label{fig:vw}}
\end{figure}

Figure~\ref{fig:vw} shows the spatial distribution of $\overline{v^\prime w^\prime}$ 
for the present cases. 
The magnitude of $\overline{v^\prime w^\prime}$ is strongest for  $h=0$ and a switch of sign  above the rough surface stripe can be seen for $h=1.70\bar{k}$. 
The distribution for $h=1.70\bar{k}$ corresponds to the one found over strip-type roughness \citep{Chung2018similarity} while the distribution for  $h=0$  is in good agreement with the studies of ridge-type roughness \citep{hwang2018secondary,Vanderwel2019}.
The opposing signs of $\overline{v^\prime w^\prime}$ for $h=0$ and $h=1.70\bar{k}$ around the smooth-rough transition location can be directly related to the different deflection of spanwise velocity fluctuations. 
The sign of the generated  correlation between $v^\prime$ and $w^\prime$ differs above the rough to smooth transition depending on whether the roughness or the smooth part of the wall forms the protruding surface.  

In case of the recessed roughness ($h=1.70\bar{k}$) the deflection on the protruding smooth surface part supports the $\overline{v^\prime w^\prime}$ distribution found on non-elevated surfaces with increased drag. In consequence, only one pair of secondary vortices is present which coincides with the one found for strip-type roughness.
For the protruding roughness ($h=0$), on the other hand, the $\overline{v^\prime w^\prime}$ distribution, opposes the one for strip-type roughness. 
In the present case, this influence of the \textcolor{black}{local mean surface} elevation dominates the secondary flow formation and thus yields a different rotational direction than for $h=1.70\bar{k}$.
For the case with $h=0.97\bar{k}$, where the melt-down height of the roughness is the same as the smooth surface height, the $\overline{v^\prime w^\prime}$ distribution appears to be dominated by the protruding parts of the surface roughness for the present geometry. 
At the same time its influence on the secondary flow formation appears to be weak.  
Overall, the present results suggest that the variation of rotational direction for different roughness heights is strongly related to the difference in the introduced wall-normal deflections of spanwise velocity fluctuations.

\section{Conclusions}


DNS of turbulent flow over alternating, streamwise-elongated, rough and smooth stripes are presented. 
The roughness is fully resolved numerically by means of an immersed boundary method. 
While the \textcolor{black} {statistical properties of the} roughness texture as well as the width and spacing of the rough areas are kept constant, 
the elevation of the smooth wall is systematically varied.
This set-up allows identifying the relevance of protruding or recessed roughness for the secondary flow formation. 
In addition, it couples the effect of lateral drag variation and \textcolor{black}{relative roughness elevation} whose effects on the secondary flow formation have been mostly studied separately in literature up to now (strip-type roughness vs. ridge-type roughness). 

The obtained results reveal opposite rotational directions for the same \textcolor{black}{ type of} roughness topography depending on whether it is introduced as  protruding  roughness ($h=0$) or  recessed  roughness ($h=1.70\bar{k}$).
While the drag on the rough surface stripes is always larger than on the smooth stripes, the secondary flow induces low speed regions above the protruding roughness  in contrast to  high speed regions above the recessed roughness.
Thus, the secondary flow  caused by protruding roughness stripes is similar to the behaviour previously reported for ridge-type roughness 
while that for a recessed roughness resembles the one over strip-type roughness. 
An intermediate case in which the mean roughness height is identical to the smooth wall position ($h=0.97\bar{k}$)  resembles the one for strip-type roughness to some \textcolor{black}{extent} and
produces significantly weaker secondary motion than the two other cases. The global drag on all heterogeneous rough surfaces is significantly larger than the area-weighted superposition of the smooth and rough  values would suggest. This drag increase appears to be related to the strength of the secondary motion. 


Since the areas with high turbulent kinetic energy and total shear stress are concentrated above the rough stripes for all investigated cases,  these quantities cannot be directly related to the observed switch in rotational direction. 
The turbulence property that is found to be related to this switch is the $\overline{v^\prime w^\prime}$ Reynolds stress component. 
This quantity, 
\textcolor{black}{ which is related to the transport of turbulent kinetic energy \citep{hwang2018secondary} and} whose spatial gradients occur in the mean momentum budget for $\overline{v}$
and $\overline{w}$ \citep{Stroh2016}, switches sign in agreement with the rotational direction of the secondary motion. 
This sign switch is related to the relative roughness height through the different deflections that spanwise velocity fluctuations experience for protruding or recessed roughness. 
For recessed roughness the generated  $\overline{v^\prime w^\prime}$-distribution is similar to the one for idealized strip-type roughness. 
Therefore, an elevated smooth surface part potentially enhances the strength of the secondary motion. 
\textcolor{black}{For protruding roughness} the deflections at the rough-smooth transition are such that a competing mechanism for the secondary flow formation is generated. 
With increasing roughness height this effect is increasingly dominant and can generate a switch of the large scale rotational direction of the secondary motion. 
Thus the relative roughness height is identified as a key quantity for the rotational direction of secondary flow over spanwise heterogeneous roughness.
\textcolor{black}{We note that this effect might be less pronounced in high Reynolds number flows for which the ratio $k_{max} / \delta $ can be significantly smaller. This issue should be addressed in future experimental studies.   
}



Finally, the present results suggest that it could be possible to control strength and rotational direction of the secondary motions above inhomogeneous rough surfaces through the relative roughness elevation.
Such a control option is highly interesting since the induced secondary motions indicate a significant global drag increase irrespective of their rotational direction while a minimum of drag increase is expected for the transition between protruding and recessed roughness. It remains to be tested in future studies which minimal drag can be achieved for inhomogeneous rough surfaces through minimization of the secondary motions.

\section*{Acknowledgements}
\textcolor{black}{We greatly appreciate  helpful discussions with  Ramis \"Orl\"u and thank the reviewers for insightful questions.}
Support by the German
Research Foundation (DFG) under Collaborative Research Centre SFB/
TRR150 project B02 is greatly acknowledged.
This work was performed on the computational resources bwUniCluster, ForHLR Phase I \& II, funded by the Ministry of Science, Research and the Arts Baden-W\"{u}rttemberg, and DFG within the framework program bwHPC.
The data presented in the manuscript are openly available in the KITopen repository at \url{http://dx.doi.org/10.5445/IR/1000100142}.

\section*{Declaration of Interests}
The authors report no conflict of interest.

\bibliographystyle{jfm}
\bibliography{lit.bib}

\end{document}